\documentclass[preprint2]{aastex}

\slugcomment{Accepted for publication in ApJ}

\begin{document}

\title
{On the Energy  Required to Eject Processed Matter from Galaxies}

\author{Sergiy Silich\altaffilmark{1,2} and Guillermo Tenorio-Tagle
\altaffilmark{1}}
\altaffiltext{1}{Instituto Nacional de Astrofisica Optica y Electronica, 
AP 51, 72000 Puebla, Mexico}
\altaffiltext{2}{Main Astronomical Observatory National Academy of Sciences 
of Ukraine, 03680 Kiyv-127, Golosiiv, Ukraine}

\begin{abstract} We evaluate the minimum energy input rate that starbursts 
require for expelling their newly processed matter from their host galaxies. 
Special attention is given to the  pressure caused by the environment in
which a galaxy is situated, as well as to the intrinsic rotation of the 
gaseous component. We account for these factors and for a massive dark matter 
distribution, and develop a self-consistent solution for the interstellar 
matter gas distribution. Our results are in excellent agreement with the 
results of Mac Low \& Ferrara (1999) for galaxies with a flattened disk-like 
ISM density distribution and a low intergalactic gas pressure 
($P_{IGM}/k$ $\leq $ 1 cm$^{-3}$ K). However, our solution also requires a 
much larger energy input rate threshold when one takes into consideration 
both a larger intergalactic pressure and the possible existence of a 
low-density, non-rotating,  extended gaseous halo component. 
\end{abstract}


\section {Introduction}
 
The observational evidence of  powerful starbursts in dwarf galaxies  has led 
to the idea that, due to their rather shallow gravitational potential, 
supernova (SN) products and, even the whole of the interstellar medium (ISM), 
may be easily ejected from the host dwarf systems, causing the contamination 
of  the intra-cluster medium. This issue has been addressed in several papers.
De Young \& Gallagher (1990) and De Young \& Heckman (1994) concluded that 
typically 1/6 of the shock-wave's swept-up mass may escape its host galaxy, 
and coherent SN explosions may even totally destroy a galaxy's ISM. More 
recently, however, Silich \& Tenorio-Tagle (1998) and  D`Ercole \& Brighenti 
(1999) have shown that within a dark matter (DM) and a low-density extended 
gaseous halo, a galactic wind does not easily develop, making the  removal 
of the galaxy's ISM very inefficient. Recently, Mac Low and Ferrara 
(1999, hereafter MF) have used several observational 
relations to define the galactic dark matter component by means of the central
DM density ($\rho_c$), its maximum extent ($R_h$) and its characteristic scale
($R_c$). They made quantitative conclusions regarding the energetics required 
for the ejection of metals and of the ISM ("blow-away")  from galaxies with 
an ISM mass in the range (M$_{ISM}$ = 10$^6$ - 10$^{10}$ M$_{\odot}$).

In none of the above mentioned studies, however, have the roles of the 
extragalactic environment nor of the rotation of the interstellar gas been 
examined in full detail. Meanwhile, Babul \& Rees (1992) have indicated that
the extragalactic medium pressure ($P_{IGM}/k$) may vary between 10$^{-1}$ and 
10$^4$ cm$^{-3}$ K, making  it one of the key parameters in the 
galactic-extragalactic medium connection.

In the present paper we reexamine  the efficiency of starbursts in  ejecting 
their newly produced metals, and in ejecting their whole ISM, from dwarf and 
normal galaxies, taking into  account  new boundary and initial conditions. 
These  follow from the pressure equilibrium condition at the 
galactic-extragalactic boundary, and the relative ability of rotation and 
interstellar gas pressure to balance different fractions of the galaxy's
gravity. In particular, we show that in a self-consistent model, the ISM's gas 
velocity dispersion and its density at the galaxy center 
are no longer free parameters, but are defined by the total ISM mass 
($M_{ISM}$), the intergalactic pressure ($P_{IGM}$), and the fraction of the 
radial component of gravity  balanced by pressure gradients. It follows from 
our considerations that the models of MF which have a 
low velocity dispersion and a flattened ISM disk-like density distribution, 
represent a limiting case in the more general scenario of starburst dwarf 
galaxy evolution. 
 
Section 2 presents the derivation of our model galaxy. Section 3 is 
devoted to a search for the threshold energy input rate required to expel 
either metals or the ISM from galaxies with a total ISM mass in the
range of 10$^6$ to 10$^9$ M$_{\odot}$. Section 4 is a discussion of our 
results.   

\section {The Galaxy Model}

Our initial model (Silich \& Tenorio-Tagle, 1998; hereafter referred to as 
Paper I) included stars, dark matter, and several isothermal interstellar gas 
components related to the ionized, neutral and  molecular ISM phases. It was 
in fact an extension to the model developed by Morita (1982), Tomisaka \& 
Ikeuchi (1988), Tomisaka \& Bregman (1993) and Suchkov et al. (1994). To 
be consistent with the approach of MF we have 
simplified our model by using only the gravitational field imposed by the
dark matter (DM) component and using only one of the  ISM phases with an 
effective velocity dispersion C$_{ISM}$. 

The spheroidal DM mass distribution can be approximated by  
\begin{equation}
      \label{eq.1}
\rho_{DM}(r) = \frac{\rho_c}{1 + (r/R_c)^2}.
\end{equation}
The DM halo radius ($R_h$), characteristic scale height ($R_c$) and 
central density ($\rho_c$) are given by MF:
\begin{eqnarray}
      \label{eq.2}
      & & \hspace{-1.2cm}
R_h = 0.016 \left(\frac{M_{DM}}{M_{\odot}}\right)^{1/3} H^{-2/3} kpc,
      \\[0.2cm]
      \label{eq.3}
      & & \hspace{-1.2cm}
R_c = 0.89 \times 10^{-5}\left(\frac{M_{DM}}{M_{\odot}}\right)^{1/2}
      H^{1/2} kpc, 
      \\[0.2cm]
     \label{eq.4}
      & & \hspace{-1.2cm}
\rho_c = 6.3 \times 10^{10}\left(\frac{M_{DM}}{M_{\odot}}\right)^{-1/3} 
               H^{-1/3} M_{\odot} \, kpc^{-3},
\end{eqnarray}
where $H$ is the Hubble constant in 100 km s$^{-1}$units. We adopt 
$H$=0.65 throughout the paper. 

On the other hand, the interstellar gas density and pressure  distributions  
are then defined by (see paper I) 
\begin{eqnarray}
      \label{eq.5}
      & & \hspace{-0.5cm}
\rho_{ISM} = \rho_0 \exp{\left[\frac{3}{2}
\left(\frac{V_h}{C_{ISM}}\right)^2\chi \right]},
      \\[0.2cm]
      \label{eq.9}
      & & \hspace{-0.5cm}
P_{ISM} = \frac{1}{3} \rho_{ISM} C_{ISM}^2.
\end{eqnarray}
where $\rho_0$ is the ISM density at the galactic center, 
$V_h = \sqrt{\frac{2GM_{DM}}{R_h}}$ is the escape velocity at the DM boundary 
$R_h$, $G$ is the gravitational constant and $M_{DM}$ is the DM mass, which
is given by
\begin{eqnarray}
      \label{eq.6}
      & & \hspace{-0.5cm}  \nonumber
M_{DM} = 4 \pi \int_0^{R_h} \rho(\omega)\omega^2{\rm d}\omega = 
      \\[0.2cm] 
      & & \hspace{-0.5cm}
4\pi\rho_{c}R_c^3 \left[y_h - \arctan{(y_h)}\right].
\end{eqnarray} 
The function $\chi$ is 
\begin{equation}
      \label{eq.7}
\chi = F(\omega) - \epsilon^2 F(r) - (1-\epsilon^2) F(0),
\end{equation}
where $r = \sqrt{(x^2 + y^2)}$ and $\omega = \sqrt{(x^2 + y^2 + z^2)}$ are
the cylindrical and spherical radii, respectively, and the dimensionless radii
$y_{\omega} = \omega/R_c$ and $y_h = R_h/R_c$. The factor $(1-\epsilon^2)$ is 
the fraction of the radial component of gravity that is balanced by pressure 
gradients (Tomisaka \& Ikeuchi, 1988). The case $\epsilon = 1$ implies 
a full balance between the radial component of gravity and the circular 
rotation of the galaxy, whereas $\epsilon = 0$ implies spherically symmetric 
systems without rotation. The function F was defined in paper I and, in the 
case of a potential exerted by DM only, it becomes:
\begin{eqnarray}
\label{eq.8} 
& & \hspace{-0.7cm}
F(\omega) = 1 + \frac{R_h}{R_c} \frac{4\pi\rho_{c}R_c^3}{M_{DM}} \times
\left[\frac{1}{2}\left(\ln{(1+y_h^2)} - 
\right. \right.
\nonumber \\[0.2cm] & & \hspace{-0.7cm}
\left. \left.
\ln{(1+y_{\omega}^2)}\right) +
\frac{\arctan{(y_h)}}{y_h} - \frac{\arctan{(y_{\omega})}}{y_{\omega}}\right],
\nonumber \\[0.2cm] & & \hspace{-0.7cm}
\, for \quad \omega \le R_h,
\\[0.2cm] & & \hspace{-0.5cm}
F(\omega) = \frac{R_h}{\omega}, \quad for \quad \omega > R_h, 
\end{eqnarray} 

The DM mass is related to the visible mass by (MF)
\begin{equation} 
\label{eq.6a} 
M_{DM} = 3.47 \times10^8
\left(\frac{M_{ISM}}{10^7M_{\odot}}\right)^{0.71} \, M_{\odot}.
\end{equation}

One can also include a transition from a rotating disk to a 
non-rotating spherical halo by expressing the $\epsilon$ in 
equation (\ref{eq.7}) as 
\begin{equation}
      \label{eq.10}
\epsilon = \epsilon_c / \exp{[(z/H_z)^2 + (r/H_{r})^2]},
\end{equation}
where $\epsilon_c$ is the value of $\epsilon$ at galactic center. This 
prevents the funnel-like density distribution that develops around the 
symmetry axis (see Suchkov at al. 1994). However, to be consistent with the 
considerations of MF, we have set the scale-lengths H$_z$ and H$_r$ to be
much greater than or comparable to the DM halo radius.
 
Following MF we assume that the interstellar gas extends out to a cutoff 
radius $R_{ISM}$ given by
\begin{equation}
      \label{eq.11}
R_{ISM} = 3 \left(\frac{M_{ISM}}{10^7M_{\odot}}\right)^{0.338} kpc.
\end{equation}
However, we further assume that the galaxy is stable and that the ISM is 
in pressure balance with the intercluster gas at the galaxy outer
boundary ($P_{ISM}$ = $P_{IGM}$). This defines the ISM gas density
at the galaxy outer edge to be
\begin{equation}
      \label{eq.12}
\rho_G = \frac{3 P_{IGM}}{C_{ISM}^2}.
\end{equation}

Thus, given the total mass of the ISM ($M_{ISM}$) and the pressure of the 
intergalactic medium ($P_{IGM}$), only $\rho_0$ and $C_{ISM}$ must be
found for fully defining the ISM mass distribution. This can be achieved by 
solving a system of two nonlinear equations 
\begin{eqnarray}
      \label{eq.13}
      & & \hspace{-0.5cm}
4 \pi \int_{0}^{R_h}\int_{0}^{R_h} \rho_{ISM}(r,z) \, r \, {\rm d}r \, 
{\rm d}z = M_{ISM}
      \\[0.2cm]
      \label{eq.14}
      & & \hspace{-0.5cm}
P_{ISM}(R_{ISM},0) = P_{IGM},
\end{eqnarray}
where $\rho_{ISM}(r,z)$ and $P_{ISM}(r,z)$ are defined by the equations
(\ref{eq.5}) and (\ref{eq.9}). Note that $\rho_{ISM}(r,z)$ should be
set equal to zero whenever its calculated value falls below $\rho_G$.
The equations can be solved by an iterative method of high accuracy. 
Thus, in a self-consistent model, the ISM gas velocity dispersion and
gas density at the galaxy's center are no longer free parameters, but are 
determined  by the total ISM mass $M_{ISM}$, the extragalactic pressure 
$P_{IGM}$, and the fraction of the radial component of gravity  balanced by
pressure gradients. 

Figure 1 presents as an example, the ISM density distribution for 
$M_{ISM}=10^8$ M$_{\odot}$ models with different fractions ($\epsilon^2$) of 
the radial component of  gravity balanced by the centrifugal forces. The 
figure shows the  smooth transition from a flattened disk-like system  
dominated by fast rotation (Panel a) to a spherical, non-rotating galaxy 
(Panel c).

\section {Energy requirements}

Once a nuclear coeval starburst has all its stars on the main sequence, it 
begins to build a superbubble. In a constant density medium, the shock enters 
a phase of continuous deceleration from the very beginning of the superbubble 
evolution. However, if the starburst is placed in a plane with a stratified 
decreasing gas density, it will expand furthest in the 
direction of least resistance, deforming the originally round superbubble 
into an elongated remnant along the symmetry axis. In the case of an exponentially-decreasing 
density distribution, the remnant will then 
break out  if the leading shock reaches 
a couple of galactic scale heights ($H_e$) while maintaining a supersonic 
speed, $V_S$ (Kompaneets 1960, Koo \& McKee 1992), while in a Gaussian atmosphere 
the shock acceleration starts once it reaches a galactic scale-height $H_g$ (Koo \& McKee 1990). 
As $V_S = (P_{bubble}/\rho_Z)^{0.5}$, where $P_{bubble}$ is the interior 
pressure, and  $\rho_Z$ is the ISM density that decreases sharply along 
the $Z$ axis, the shock is forced to accelerate. Once in the acceleration 
phase, neither the external gas pressure nor the galaxy's gravitational 
potential can restrain the shock, which ultimately will acquire speeds larger 
than the galactic escape velocity ($V_{esc}$). This occurs even if 
$V_s < V_{esc}$ prior to the acceleration phase that begins at $Z \sim$ the 
gaseous scale-height.
It is also well known that the swept up matter immediately behind the 
accelerating shock, also accelerates, producing the onset of Rayleigh-Taylor 
instabilities that fragment the swept-up matter shell. The remnant is then 
able to ''blowout'', driving  its high pressure gas between the fragments 
into the halo of the host galaxy, where it once again will push 
the shock (see Tenorio-Tagle \& Bodenheimer 1988, and references therein).  

The minimum starburst energy input rate ($L_{cr}$), that leads to
breakout, continuous acceleration, and expulsion of matter from galaxies
into the IGM, either from a Gaussian or an exponential atmosphere 
was studied by Koo \& McKee (1992). The authors used the 
threshold luminosity criterion
($L_{breakout}$)  which assures that the
shock will reach one scale height $H$ while maintaining a supersonic
speed: 
\begin{equation}
      \label{eq.22} L_{breakout} = 17.9 \rho_0 H^2 C_{ISM}^3 \, ergs \,
s^{-1} \end{equation}
For an exponential atmosphere, Koo \& McKee (1992) demand an
$L_{cr} \geq 3L_{breakout}$. In a Gaussian density distribution the
acceleration phase starts earlier, and thus we shall assume 
$L_{cr} \geq L_{breakout}$. Note, that in the flattened disk-like
configurations, the characteristic scale height $H$ may exceed the
thickness of the galaxy's disk $Z_{ISM}$, requiring that $H$ in the
equation (\ref{eq.22}) be replaced by $Z_{ISM}$.

\subsection{The Density Distribution along the Axis of Symmetry}

Applying the breakout criterion to our model galaxies required 
accurate fits to the resultant density distributions (see Section 2) along 
the symmetry axis.

\subsubsection{Disk-like density distributions}

In the case of flattened disk-like distributions, one can show that near the 
galaxy's plane the resultant gas density distribution  along the symmetry 
axis is fitted well by a Gaussian function:
\begin{equation}
      \label{eq.15}
\rho_{ISM} = \rho_0 \exp[-(Z/H_g)^2], 
\end{equation}
with the characteristic scale height
\begin{equation}
      \label{eq.16}
H_g = \sqrt{\frac{C_{ISM}^2}{2 \pi G \rho_c}}.
\end{equation}
Further away from the galaxy's plane, the density distribution can be well 
approximated by an exponential function 
\begin{equation}
      \label{eq.17}
\rho_{ISM} = \rho_G \exp[(Z_{ISM} - Z) / H_e],
\end{equation}
where $Z_{ISM}$ is the disk thickness along the Z axis. As the match to the 
density distribution requires two different functions, we have restricted 
the parameter $H_e$ by the condition that the density gradient, as given by 
the functions (\ref{eq.15}) and (\ref{eq.17}), must be smooth and continuous 
at the crossing point ($Z_f$). This condition can be represented by the 
following two equations:
\begin{eqnarray}
      \label{eq.18}
      & & \hspace{-0.5cm} \nonumber
log \rho_0 - \left(\frac{Z_f}{H_g}\right)^2 log(e) =
      \\[0.2cm]
      & & \hspace{-0.5cm}
log \rho_G + \frac{Z_{ISM} - Z_f}{H_e} log(e),
      \\[0.2cm]
      \label{eq.19}
      & & \hspace{-0.5cm}
\frac{2 Z_f}{H_g^2} log(e) = \frac{1}{H_e} log(e).
\end{eqnarray}
Solving for H$_e$ and Z$_f$ gives:
\begin{eqnarray}
      \label{eq.20}
      & & \hspace{-0.5cm}
H_e = \frac{1}{2} \frac{Z_{ISM}^2}{Z_f},
      \\[0.2cm]
      & & \hspace{-0.5cm} \nonumber
Z_f = Z_{ISM} \times
      \\[0.2cm]
      \label{eq.21}
      & & \hspace{-0.5cm}
\left[1 - \sqrt{1 - \left(\frac{H_g}{Z_{ISM}}\right)^2
      \frac{log(\rho_0/\rho_G)}{log(e)}}\right],
\end{eqnarray}
which uniquely define the intersection point  $Z_f$ and the fitting parameter 
$H_e$ required in estimating the minimum energy input rate for breakout.

The ISM distribution along the Z axis for different disk-like galaxies and 
different  extragalactic gas pressures (solid lines) and their analytic fit 
(dashed lines) are shown in the Figures 2 a and b. 

\subsubsection{Spherical density distributions}

The ISM density distribution for different  galaxies under the assumption of 
$\epsilon$ = 0, and different  extragalactic gas pressures are shown in  
Figures 2 c and d. In these cases the density distribution far from the 
galaxy's plane cannot be fitted by a Gaussian or an an exponential function. 
Thus to  estimate the minimum energy input rate required to  eject the metals 
from these galaxies we have only used the numerical scheme of paper I. In all 
cases we have assumed a constant mechanical energy input rate during the 
starburst supernova phase (4 $\times 10^{7}$ yr).

\subsubsection{The energy requirements}

Our galaxy models  are summarized in Table 1. The various models 
are labeled with several indexes that represent the 
logarithm of the ISM mass considered, $\epsilon$ values that range between 0 
and 0.9, and intergalactic gas pressure ($P_{IGM}/k$) values that are assumed 
to be either 1 or 100 cm$^{-3}$ K. In the Table, column 1 gives the model 
identification. Columns 2 and 3 list the DM and ISM masses, column 4 gives 
the ISM cutoff radius, column 5 the intra-cluster gas pressure ($P_{IGM}$), 
and columns 6 and 7 show the interstellar gas velocity dispersion and escape 
velocity at the galaxy boundary $R_{ISM}$. The central gas number density is 
shown in column 8 and column 9 shows the ISM extent along Z axis ($Z_{ISM}$).

Figure 3 shows our energy estimates  
resultant from the numerical integration of the 
hydrodynamic equations for a coeval starburst with a constant energy input 
rate during the first 4 $\times 10^7$ yr of the evolution,  and the initial 
density distributions shown in Figure 2 a - d. 
Figure 3 considers a range of galactic ISMs 
(from 10$^6$ to 10$^9$ M$_{\odot}$) and extreme values of $\epsilon$ 
(= 0.9 for flattened disk-like density distributions and = 0 for spherical 
galaxies without rotation) and also for a range of values of $P_{IGM}/k$ 
(equal to 1 and 100 cm$^{-3}$ K). Our results for disk-like gas distributions
are in excellent agreement with the findings  of MF, particularly for the 
case of $P_{IGM}/k$ = 1 cm$^{-3}$ K. Note, however, that the threshold energy 
input rate in our models depends on the  intergalactic gas pressure, 
$P_{IGM}$, and becomes rapidly larger for smaller $M_{ISM}$ and 
larger $P_{IGM}$. 

In the case of the flattened disk-like configurations ($\epsilon$ = 0.9; see 
Table 1), our numerical experiments indicate that the shock acceleration 
phase starts after 1.2 - 1.4 $H_g$, and thus it is the Gaussian rather than 
the exponential outer part of the disk-like density distributions the ones 
that define the onset of shock acceleration. Our results are in good 
agreement with the analytic criterion that ensures breakout 
(equation (17)) from the density distributions shown in 
Figure 2 a and b, particularly when the values of the Gaussian 
scale-height ($H$ = $H_g$) are used (filled symbols in Figure 3). 

In Figure 3, values below each line imply total retention, while the region 
above each line indicates the expulsion of the hot superbubble interior gas 
(the new metals) out of disk-like distributions ($\epsilon$ = 0.9), 
and of the new metals and the whole of the ISM in the spherical 
($\epsilon$ = 0) cases. 
Each line that separates the two regions  marks the 
minimum energy input rate needed from a coeval starburst to reach the outer 
boundary of a given galaxy, regardless of the time that the remnant may 
require to do so. In the case of disk-like configurations, this energy input 
rate warrants breakout and the ensuing continuous acceleration. In most of 
these cases the ejection of the hot gas into the intergalactic medium occurs 
before the starburst supernova phase is over (4 $\times 10^7$ yr). 
In galaxies with $\epsilon$ = 0, however, the evolution time considered 
largely exceeds the starburst supernova phase. This causes the  decelerating 
shell of interstellar swept up matter to reach the galaxy edge by means of the 
momentum gathered during its early evolution. Figure 4 shows the results for 
a $M_{ISM}$ = 10$^8$ M$_\odot$ galaxy under two extremes of 
intergalactic pressures ($P_{IGM}/k$ = 1 and 100 cm$^{-3}$ K) as they become 
expelled by the lower limit mechanical energy input rate displayed in 
Figure 3. Throughout its evolution, the remnant in the high pressure galaxy 
undergoes a continuous deceleration, while the low pressure system is able 
to experience the acceleration promoted by breakout. In both cases 
however, the time needed to reach the galaxy outer boundary is much larger 
than the supernova phase and thus both remnants slow down by conservation of 
their own momentum, to reach the galaxy boundary and experience total 
disruption after 100 Myr and 200 Myr of evolution in the high and low 
intergalactic pressure cases, respectively.

The right-hand axis in our summary Figure 3 indicates the mass of the coeval 
starburst able to produce the energy input rate indicated in the left-hand 
axis. This has been scaled from the synthesis models of Leitherer \& Heckman 
(1995) for a 10$^6$ M$_\odot$ starburst under the assumption of a Salpeter IMF
with a lower mass limit of 1 M$_\odot$, an upper stellar mass equal to 100 
M$_\odot$, and a metalicity $Z$ = 0.1. In this case, the log of the mechanical
energy input rate in erg s$^{-1}$ is 40.5, and scales linearly with the 
starburst total mass.

In Figure 3, we have also indicated the results of another two calculations 
where we  assumed a gravitational potential corresponding to 
a  $M_{ISM}$ = 10$^9$ M$_\odot$  galaxy. In these two models however, we have 
adopted a spherical halo with total masses of 10$^8$ M$_\odot$ 
and 10$^7$ M$_\odot$ and  an intergalactic gas pressure equal to 
$P_{IGM}/k$ = 1 cm$^{-3}$ K. The derived minimum energy input rate 
(or minimum mass of the coeval starburst required to expel the ISM)  
clearly indicates that it is the mass of the halo, instead of the  disk 
component, what determines the minimum energy input rate necessary for
ejection. Figure 3 shows that a halo with only 1/100 of the mass of the 
disk-like configuration ($M_{ISM}$ = 10$^9$ M$_\odot$) increases  the value
of the minimum energy input rate required for mass ejection  by approximately 
an order of magnitude. An  energy input rate nearly two orders of magnitude 
higher is needed if the spherical halo has a mass of 1/10 of the disk-like 
configuration.

\section{Discussion}

We have described a simple, self-consistent, approach  to build the ISM 
gas distribution of galaxies,  accounting for the gravitational potential 
exerted by a massive DM component, as well as for rotation of the ISM 
and the intergalactic gas pressure. The numerical experiments and analytical 
estimates have then showed that the final fate of the matter
ejected from a starburst region, as well as that of the  shocked ISM, 
is highly dependent on the boundary conditions. Following a continuous 
transition from a fast rotating, thin, disk-like to a spherical non-rotating 
interstellar gas distribution, we have found that the ISM to be more 
resistant to ejection than estimated in earlier papers.

We have shown (see Figure 3) that superbubbles evolving in  
galaxies that have a gaseous disk-like density distribution are likely to 
undergo the  phenomenon of breakout. This  allows them to accelerate and 
expel all of their newly produced metals, and perhaps even a small fraction 
of the interstellar medium, into intergalactic space. On the other hand,   
much larger energy input rates, or more massive coeval starbursts (up to 3 
orders of magnitude larger), are required to provoke breakout or push
a shell to the galaxy outer boundary for a spherically-symmetric ISM mass
distribution. Even low mass ($\sim$ 1\% - 10\% of the total 
ISM mass), non-rotating subsystems increase the energy requirements by more 
than an order of magnitude. This makes the low density haloes, rather than 
DM itself,  the key component in the evolution of dwarf galaxies.  

The halo properties, on the other hand, are highly dependent on the 
gravitational potential. The presence of an extended DM component makes the 
halo mass distribution more smooth and extended. However, the halo parameters 
are also dependent on the  properties of the intergalactic medium. To 
withstand a high intergalactic  pressure, the  halo has to support itself with
a high random motion, which leads to a  more homogeneous gas distribution 
compared to that found in  low pressure surroundings. 
 
Clearly the energy input rates derived here are lower limits to the amounts
required for expelling matter from a galaxy. Particularly because only one 
component of the ISM was considered and because the central densities
adopted are well below the values expected for the star forming cloud 
where the starburst originated. Our estimates thus neglect the effect
of the starburst plowing into the parental cloud  material. 
These are lower limits also because we adopted a constant energy input rate
(see Strickland \& Stevens 2000) and because our approach neglects an 
additional cooling by mass-loading process (Hartquist et al. 1986) and the 
presence of a magnetic field which also could inhibit expansion 
(Tomisaka 1998).
 
The indisputable presence of metals (in whatever abundance) in galaxies
implies that the supernova products cannot be lost in all cases. Note in 
particular that many well known disk galaxies have a high 
metal abundance and a large number of centers of star formation. Most
of these exciting star clusters are more massive than the 
$10^4$ M$_\odot$ lower limit established by MF and the current paper 
as the minimum starburst mass required to cause mass ejection 
in the case of  disk-like systems. This lower limit for disk-like galaxies 
with M$_{ISM}\leq$ 10$^9$ M$_\odot$ (see Figure 3) implies that starbursts 
even smaller than the Orion cluster would break through the galaxy outer 
boundary and eject their supernova products into the intergalactic medium.
Nevertheless, disk-like galaxies can avoid losing all their freshly 
produced metals by having a halo component, neglected in former studies, 
that acts as the  barrier to the loss of the new metals.  

The haloes, despite acting as the barrier to the loss of the new metal,
have rather low densities ($<n> \sim 10^{-3}$ cm$^{-3}$) 
and thus have a long recombination time ($t_{rec} = 1/(\alpha n_{halo}$); 
where $\alpha$ is the recombination coefficient)
that can easily exceed the life time of the  HII region ($t_{HII}$ = 10$^7$ yr)
produced by the starburst. In such a case, the haloes may remain 
undetected at radio and optical frequencies (see Tenorio-Tagle et al. 1999), 
until large volumes are collected into the expanding supershells.
Note that the continuous $\Omega$ shape that supershells present in a number 
of galaxies, while remaining attached to the central starburst, and their 
small expansion velocity (comparable or smaller than the escape velocity of 
their host galaxy) imply that the mechanical energy of the star cluster is 
plowing into a continuous as yet undetected medium. Other authors 
(see MF 1999) have argued that blowout could leave fragments of the dense 
shell behind and that these observed with poor resolution may appear to form 
a continuous shell much smaller than the true extent of the shocked region. 
Clearly, X-ray observations of these supershells will help to decide if the 
hot processed material is enclosed or not by the expanding shell.  

Thus the answer, the true limit for mass ejection from galaxies, must lie 
between the two extreme cases that we have investigated here.
Note however, that in the presence of a halo, it is the mass of the halo 
that sets the limiting energy input rate required for mass ejection, 
and not the mass of the disk-like component. This argument applies to all 
galaxies whether spirals, amorphous irregulars or dwarfs. 

We are grateful to our referee Dr. M-M. Mac Low for a speedy processing of 
our paper and the various comments that help to improve it.
We also thank  W. Wall for his careful reading of our manuscript.
GTT also thanks CONACYT for the grant 211290-5-28501E which allowed for
the completing of this study.

\clearpage

\onecolumn

\begin{figure}
\plotone{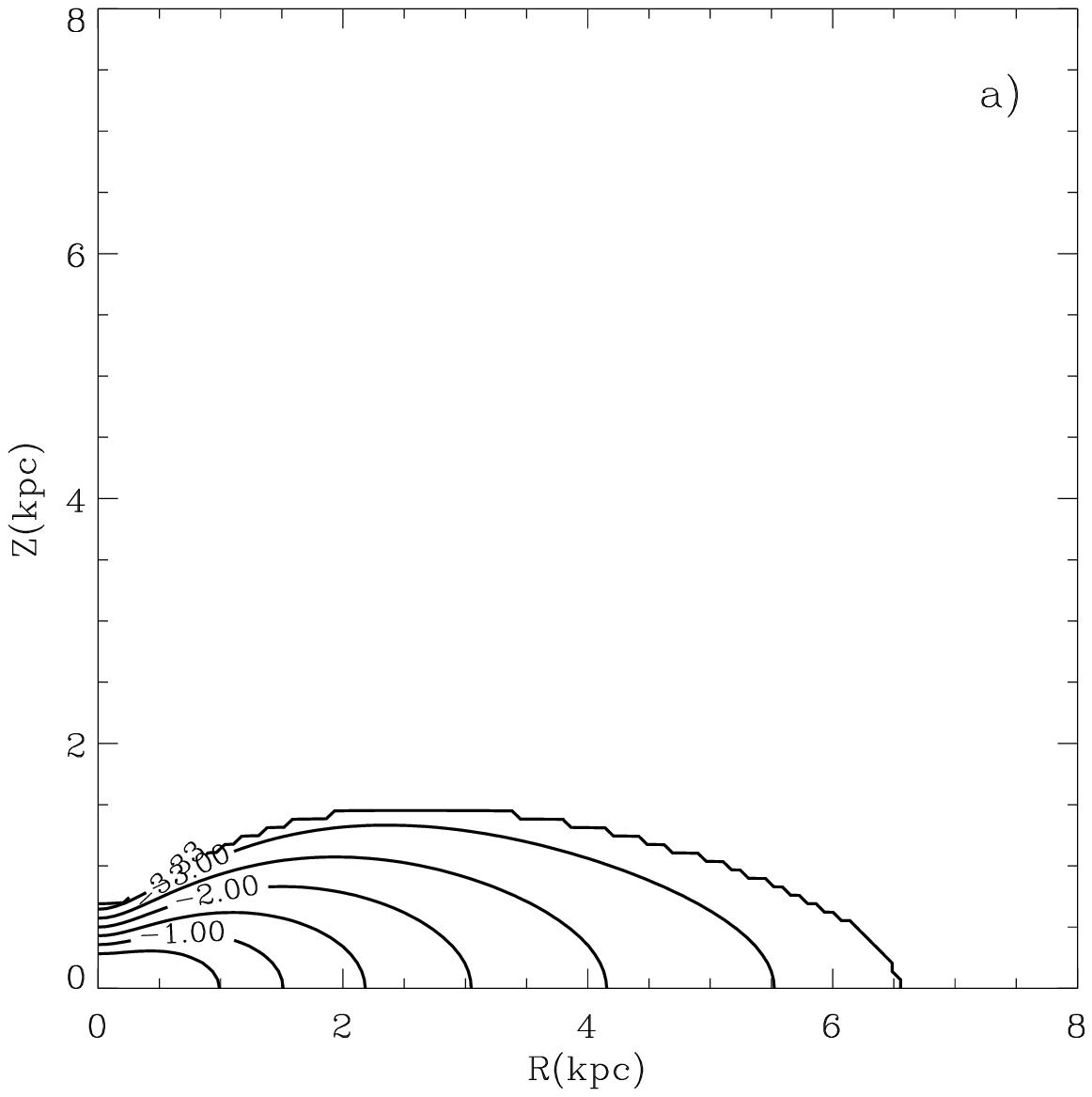}
\end{figure}

\clearpage

\begin{figure}
\plotone{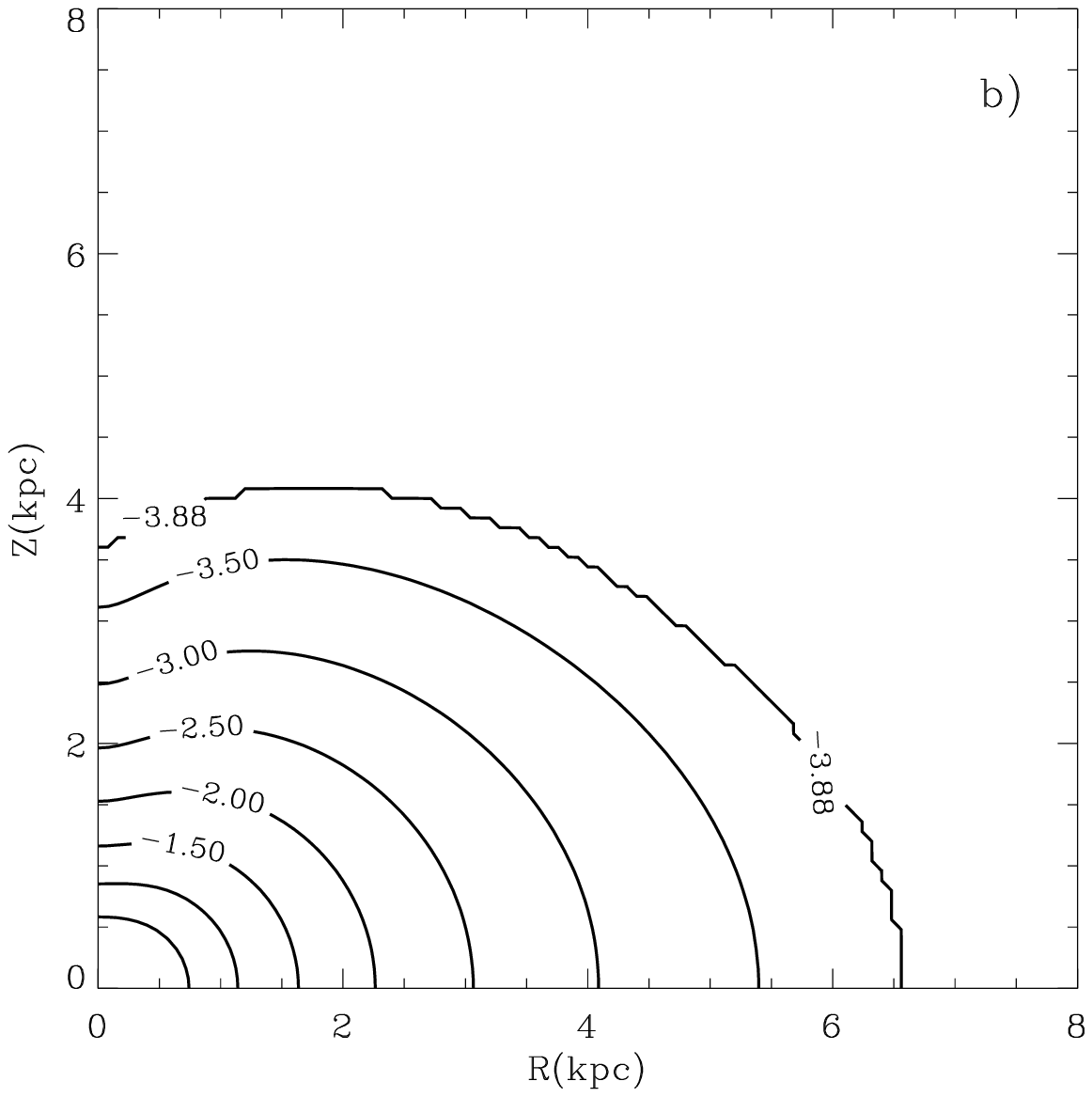}
\end{figure}

\clearpage

\begin{figure}
\plotone{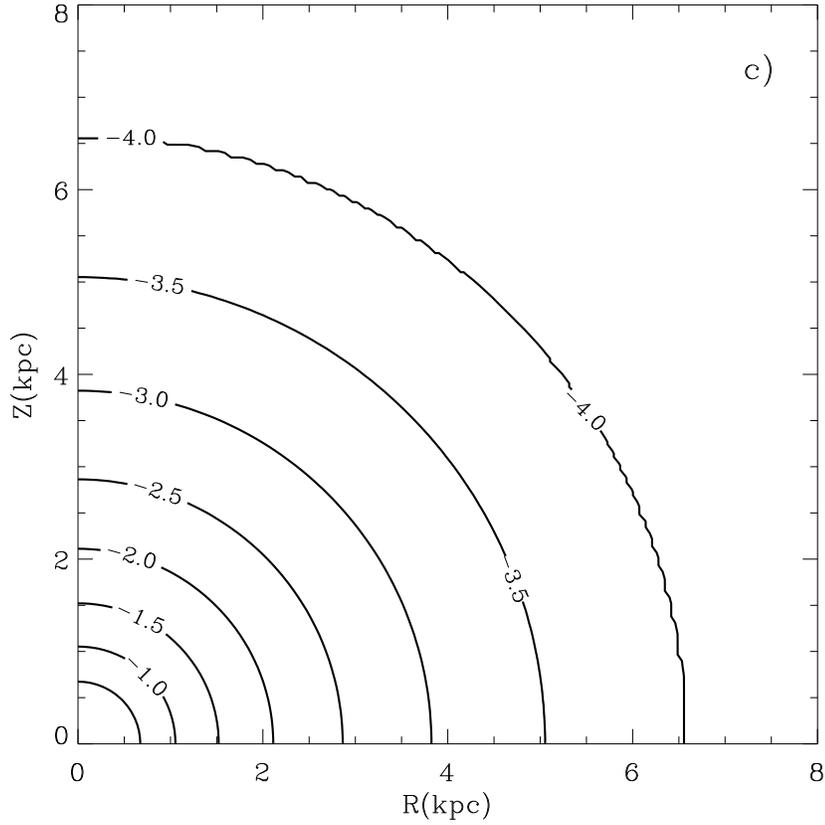}
\caption{{The model galaxy.  Logarithmic contours showing the  
density distribution for a galaxy with a $M_{ISM}=10^8$ 
M$_{\odot}$ and a $P_{IGM}/k$ = 1 cm$^{-3}$ K and several values of 
$\epsilon$ (= 0.9, 0.5, and 0.0). The latter correspond to the fraction 
$f$ =1--$\epsilon^2$ equal to 19$\%$, 75$\%$ and 100$\%$ of the radial 
component of gravity balanced by pressure gradients in panels a, b and c, 
respectively.\label{fig1}}} 
\end{figure}

\clearpage
  
\begin{figure}
\plotone{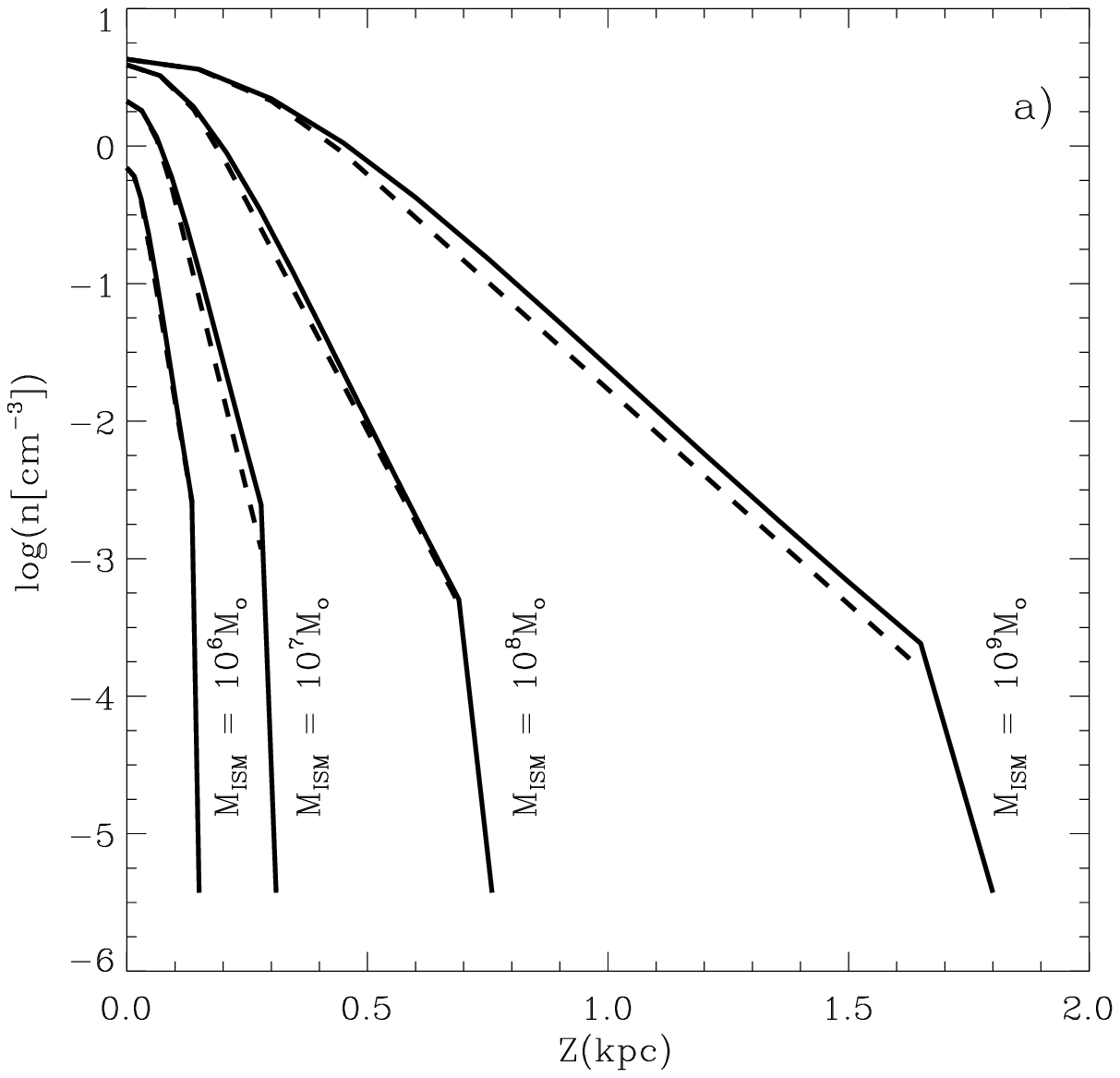}
\end{figure}

\clearpage

\begin{figure}
\plotone{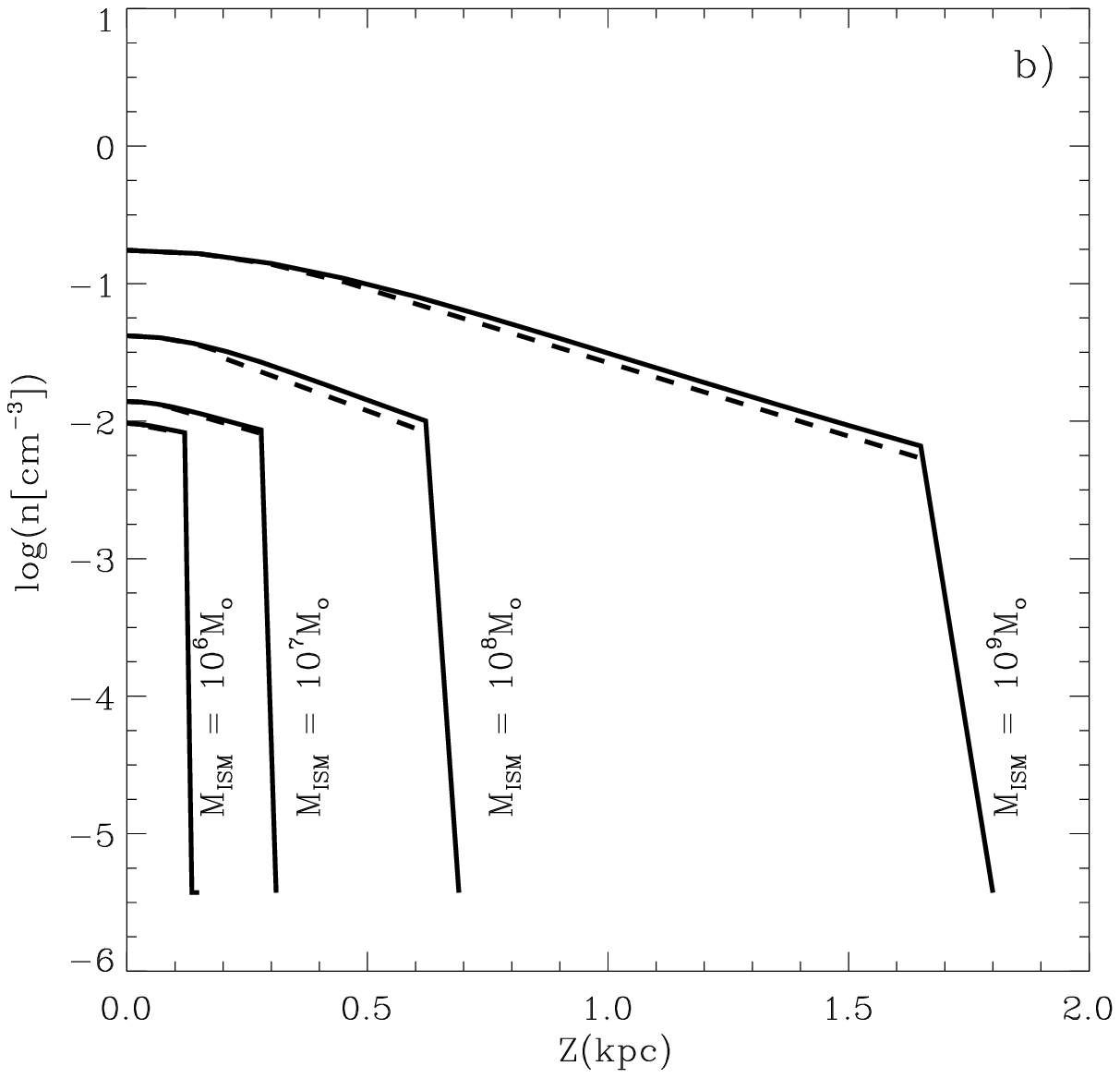}
\end{figure}

\clearpage

\begin{figure}
\plotone{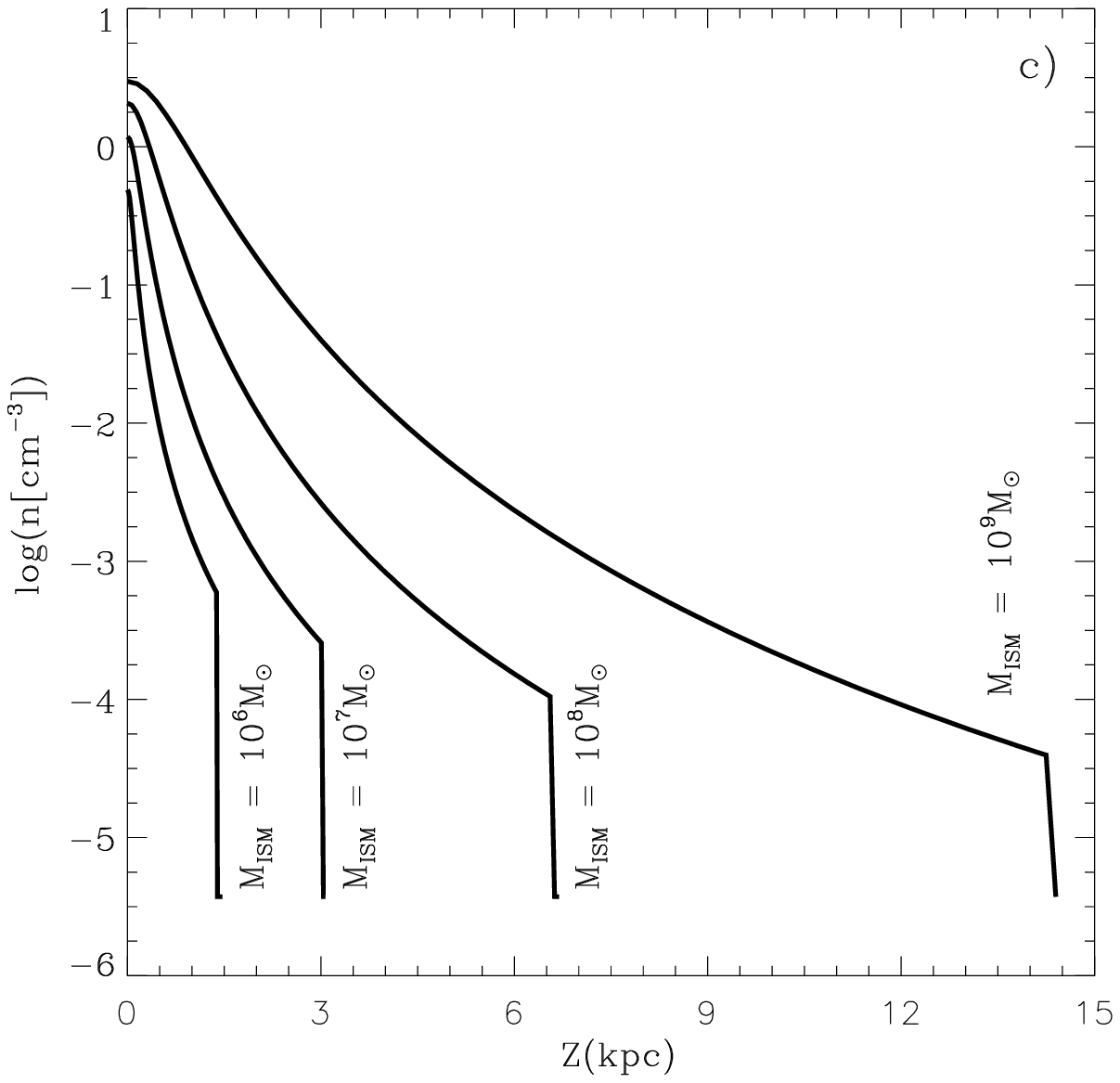}
\end{figure}

\clearpage

\begin{figure}
\plotone{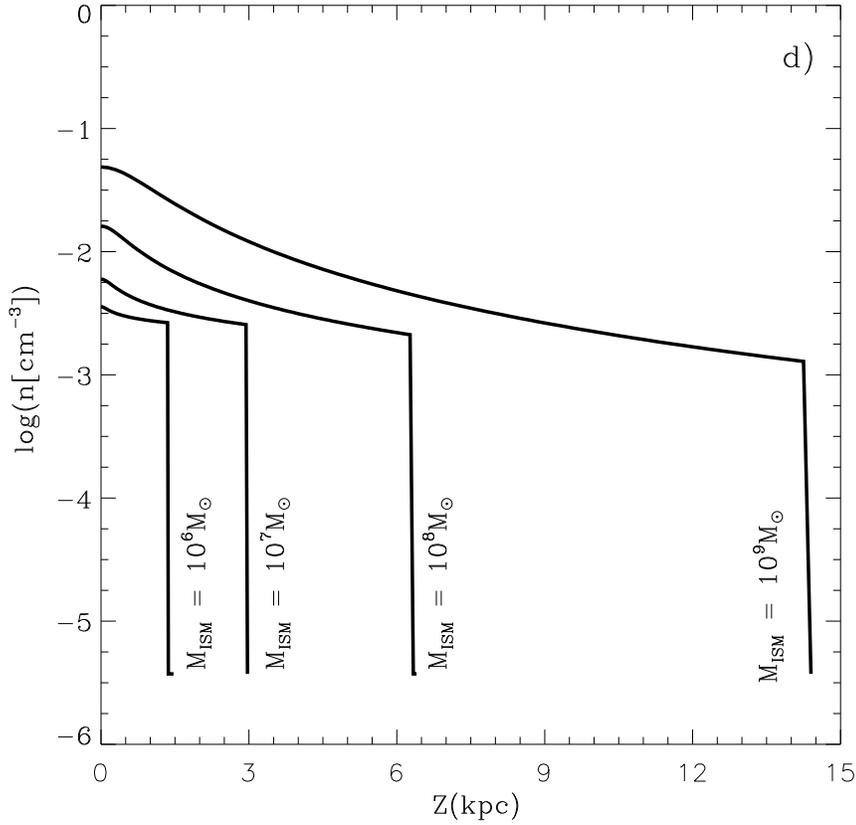}
\caption{{The density distribution. Solid lines in Panels a 
and b show the density distribution along the symmetry axis, derived for 
disk-like galaxies (see Table 1) with $\epsilon$ = 0.9 and a mass 
$M_{ISM}$ = 10$^6$ M$_{\odot}$ -- 10$^{9}$ M$_{\odot}$, and intergalactic 
pressures equal to 1 and 100 cm$^{-3}$ K, respectively. The dashed lines 
represent the corresponding fit to the density distributions. Panels c and 
d show the corresponding density distributions that resulted for galaxies 
without rotation ($\epsilon$ = 0).\label{fig2}}}  
\end{figure}

\clearpage

\begin{figure}
\plotone{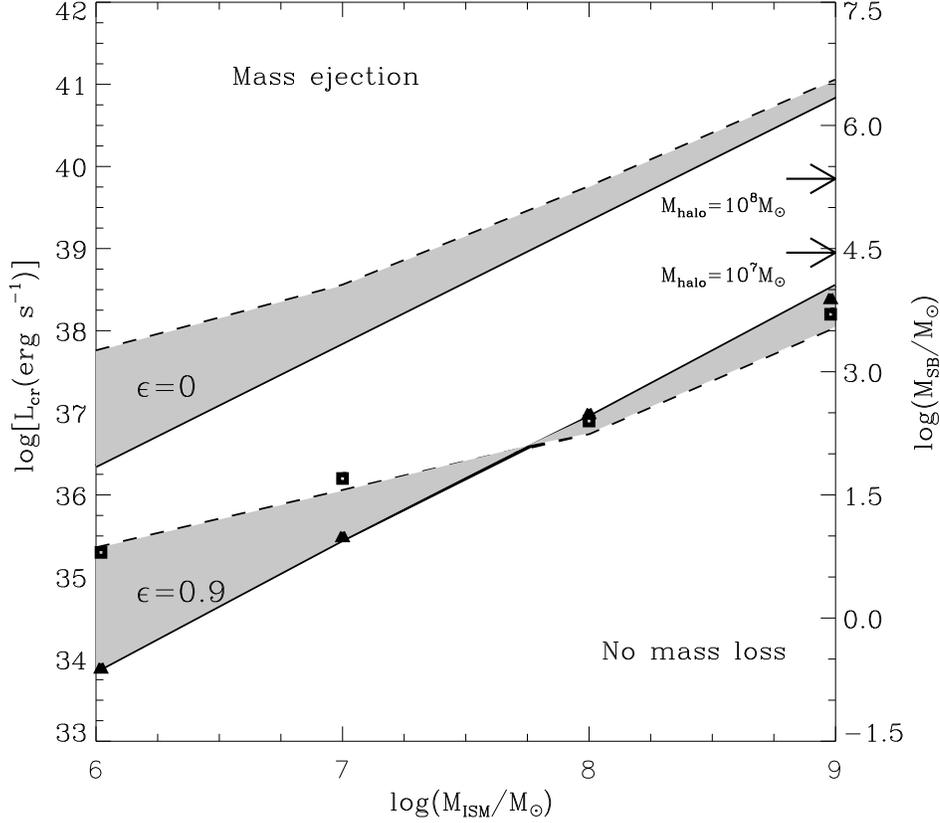}
\caption{{Energy estimates. The log of the critical mechanical 
luminosity, and of the starburst mass, required to eject matter from galaxies 
with a $M_{ISM}$ in the range 10$^6$ -- 10$^9$ M$_\odot$. The lower limit 
estimates are shown for galaxies with extreme values of $\epsilon$
(= 0 and 0.9) and for two values of the intergalactic pressure  
$P_{IGM}/k$ = 1 cm$^{-3}$ K (solid lines) and $P_{IGM}/k$ = 100 cm$^{-3}$ K 
(dashed lines). The resolution of our numerical search is 
$\Delta log L_{cr}$ = 0.1.
Each line should be considered separately as they divide the 
parameter space into two distinct regions: a region of no mass loss that is
found below the line and a region in which blowout and mass ejection 
occur that is found above the line. Also indicated on  the
right-hand axis are the energy input rates required for a remnant to 
reach the outer boundary of a halo with mass 10$^8$ M$_\odot$ and one with 
mass 10$^7$ M$_\odot$, for the case of a gravitational potential 
provided by $M_{DM}$ = 9.1$\times 10^9$ M$_\odot$ that can hold an
$M_{ISM}$ =  10$^9$ M$_\odot$.
The filled squares and triangles represent the analytical energy input rate 
estimated by means of equation (17), using  the $H_g$ values derived  for 
the central Gaussian part of the disk-like density distributions shown in 
Figure 2 a, b. \label{fig3}}}
\end{figure}

\clearpage

\begin{figure}
\plotone{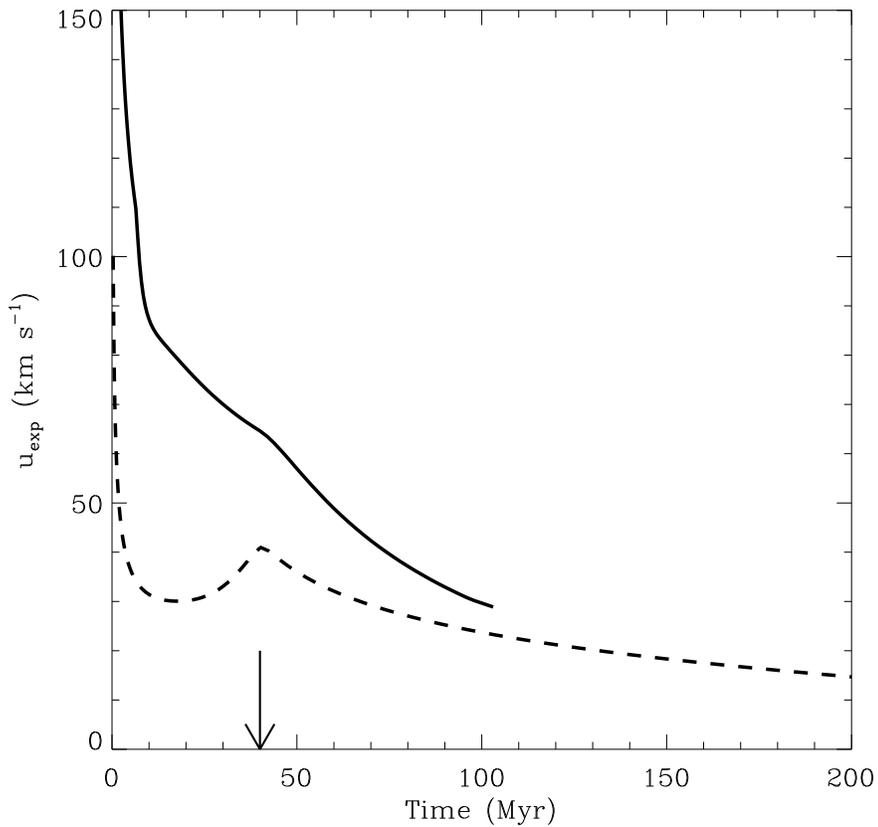}
\caption{{The maximum expansion speed measured along
the symmetry axis Z for 10$^8$M$_{\odot}$ galaxy with spherical ISM
density distribution. The solid line indicates the expansion velocity
for P$_{IGM}$/k=100, whereas the dotted line represents the shell expansion
for the P$_{IGM}$/k=1 model. The arrow indicates the end of the SB 
supernova phase. Note that the remnants require  a much longer time to
reach the outskirts of their host galaxy. \label{fig4}}}
\end{figure}

\clearpage

\begin{table}
\begin{center}
\caption{Galaxy parameters.\label{tbl1}}
\begin{tabular} {lccccccccc} 
\tableline
Model &$M_{DM}$  &$M_{ISM}$ &R$_{ISM}$ &P$_{IGM}$ & $\epsilon$ &C$_{ISM}$ 
      &V$_{esc}$ &n$_0$     &Z$_{ISM}$ \\
      & M$_{\odot}$ & M$_{\odot}$ & kpc & nT/k &  & km\,s$^{-1}$ &km\,s$^{-1}$
      & cm$^{-3}$ &pc \\
\tableline 
M600.1   &6.8$\times 10^7$ &$10^6$ &1.4 &1.0 &0.0 &5.8  &13.6 &0.5 & 1400  \\
M605.1   &6.8$\times 10^7$ &$10^6$ &1.4 &1.0 &0.5 &5.0  &13.6 &0.6 &  735  \\
M609.1   &6.8$\times 10^7$ &$10^6$ &1.4 &1.0 &0.9 &2.8  &13.6 &0.7 &135 \\
M600.100 &6.8$\times 10^7$ &$10^6$ &1.4 &100 &0.0 &27.1 &13.6 
         &3.6$\times 10^{-3}$ & 1400 \\
M605.100 &6.8$\times 10^7$ &$10^6$ &1.4 &100 &0.5 &22.7 &13.6 
         &5.2$\times 10^{-3}$ & 735 \\
M609.100 &6.8$\times 10^7$ &$10^6$ &1.4 &100 &0.9 &15.5 &13.6 
         &9.6$\times 10^{-2}$ & 120 \\
\tableline
M700.1   &3.5$\times 10^8$ &$10^7$ &3.0 &1.0 &0.0 &8.8 &22.5  &1.2 & 3000   \\
M705.1   &3.5$\times 10^8$ &$10^7$ &3.0 &1.0 &0.5 &7.7 &22.5  &1.4 & 1610   \\
M709.1   &3.5$\times 10^8$ &$10^7$ &3.0 &1.0 &0.9 &4.1 &22.5  &2.1 &279 \\
M700.100 &3.5$\times 10^8$ &$10^7$ &3.0 &100 &0.0 &27.6 &22.5 
         &6.0$\times 10^{-3}$ & 3000  \\
M705.100 &3.5$\times 10^8$ &$10^7$ &3.0 &100 &0.5 &25.4 &22.5 
         &6.5$\times 10^{-3}$ & 1645  \\
M709.100 &3.5$\times 10^8$ &$10^7$ &3.0 &100 &0.9 &15.4 &22.5 
         &1.4$\times 10^{-2}$ &279 \\
\tableline
M800.1   &1.8$\times 10^9$ &$10^8$ &6.5 &1.0 &0.0 &13.9 &36.9  &2.1 & 6500 \\
M805.1   &1.8$\times 10^9$ &$10^8$ &6.5 &1.0 &0.5 &12.1 &36.9  &2.5 & 3520  \\
M809.1   &1.8$\times 10^9$ &$10^8$ &6.5 &1.0 &0.9 &6.5  &36.9  &3.9 &690 \\
M800.100 &1.8$\times 10^9$ &$10^8$ &6.5 &100 &0.0 &30.4 &36.9 
         &1.6$\times 10^{-2}$ & 6500 \\
M805.100 &1.8$\times 10^9$ &$10^8$ &6.5 &100 &0.5 &26.4 &36.9 
         &2.2$\times 10^{-2}$ & 3600\\
M809.100 &1.8$\times 10^9$ &$10^8$ &6.5 &100 &0.9 &15.2 &36.9 
         &4.2$\times 10^{-2}$ &621 \\
\tableline
M900.1   &9.1$\times 10^9$ &$10^9$ &14.2&1.0 &0.0 &22.3 &60.4 &3.0 &14200 \\
M905.1   &9.1$\times 10^9$ &$10^9$ &14.2&1.0 &0.5 &19.5 &60.4 &3.7 & 7950  \\
M909.1   &9.1$\times 10^9$ &$10^9$ &14.2&1.0 &0.9 &11.0 &60.4 &4.3 &1650 \\
M900.100 &9.1$\times 10^9$ &$10^9$ &14.2&100 &0.0 &39.2 &60.4 
         &4.9$\times 10^{-2}$ &14200  \\
M905.100 &9.1$\times 10^9$ &$10^9$ &14.2&100 &0.5 &31.5 &60.4 
         &1.4$\times 10^{-1}$ & 7800 \\
M909.100 &9.1$\times 10^9$ &$10^9$ &14.2&100 &0.9 &19.0 &60.4
         &1.8$\times 10^{-1}$ &1650 \\
\tableline
\end{tabular}
\end{center}
\end{table}

\end{document}